\documentstyle[epsfig]{mn}
\begin{document}
\input ./BoxedEPS.tex
\newcommand{\aip}{{\small ${\cal AIPS}$}}
\newcommand{\gtsim}{\mbox{{\raisebox{-0.4ex}{$\stackrel{>}{{\scriptstyle\sim}}
$}}}}
\newcommand{\ltsim}{\mbox{{\raisebox{-0.4ex}{$\stackrel{<}{{\scriptstyle\sim}}
$}}}}
\newcommand{\s}{$\stackrel{\rm s}{.}$}
\newcommand{\h}{$^{\rm h}$}
\newcommand{\m}{$^{\rm m}$}
\newcommand{\pp}{$\stackrel{\prime\prime}{.}$}
\newcommand{\de}{$^{\circ}$}
\newcommand{\p}{$^{\prime}$}
\newcommand{\arc}{$^{\prime\prime}$}
\newcommand{\marc}{^{\prime\prime}}
\newcommand{\rs}{{\em $r_s$}}
\newcommand{\DPM}{{\em DPM}}
\newcommand{\alf}{{\displaystyle\biggl({\nu_{\rm h} \over \nu_{\rm l}}\biggr)^{\alpha}} }

\newcommand{\figstart}[1]
    { \begin{figure}[htb]
      \begin{picture}(0,#1) }
\newcommand{\figend}[4]
    { \end{picture}
      \special{#1}
      \caption[#2]{#3}
      \label{#4}
      \end{figure} }
\newcommand{\fig}[5]
    { \figstart{#1}
      \figend{#2}{#3}{#4}{#5} }
\newcommand{\bHS}{\beta_{\mbox{\scriptsize HS}}}
\newcommand{\bBF}{\beta_{\mbox{\scriptsize BF}}}
\newcommand{\nT}{\nu_{\mbox{\scriptsize T}}}
\newcommand{\et}{E_{\mbox{\scriptsize T}}}
\newcommand{\nTn}{\nu_{\mbox{\scriptsize Tn}}}
\newcommand{\nTf}{\nu_{\mbox{\scriptsize Tf}}}
\newcommand{\tn}{\tau_{x\mbox{\scriptsize n}}}
\newcommand{\tf}{\tau_{x\mbox{\scriptsize f}}}
\newcommand{\xn}{x_{\mbox{\scriptsize n}}}
\newcommand{\xf}{x_{\mbox{\scriptsize f}}}
\newcommand{\yn}{y_{\mbox{\scriptsize n}}}
\newcommand{\yf}{y_{\mbox{\scriptsize f}}}
\newcommand{\lln}{l_{\mbox{\scriptsize n}}}
\newcommand{\llf}{l_{\mbox{\scriptsize f}}}
\newcommand{\Dn}{f(\Delta_{\mbox{\scriptsize n}})}
\newcommand{\Df}{f(\Delta_{\mbox{\scriptsize f}})}
\newcommand{\B}{\mbox{$B$}}
\newcommand{\Bo}{\mbox{$B$}_{0}}

\SetEPSFDirectory{./}
\SetRokickiEPSFSpecial
\HideDisplacementBoxes

\title
[HST imaging survey of sub-mJy star-forming
galaxies I]
{Hubble Space Telescope imaging survey of sub-mJy star-forming
galaxies I: morphologies at $z\sim0.2$}

\author[Serjeant, Mobasher, Gruppioni, Oliver]{
Stephen Serjeant$^1$, Bahram Mobasher$^{1,2,3}$,
Carlotta Gruppioni$^{1,4}$, Seb Oliver$^{1,5}$\\
$^1$Astrophysics Group, Blackett Laboratory, Imperial College,
Prince Consort Road,London SW7 2BZ, UK\\
$^2$Space Telescope Science Institute, 3700 San Martin Drive,
Baltimore, MD 21218, USA\\
$^3$Affiliated with the Astrophysics Division, European Space Agency\\
$^4$Osservatorio Astronomico di Bologna, via Ranzani 1,
40127 Bologna, ITALY\\
$^5$Astronomy Centre, CPES, University of Sussex,
Falmer, Brighton BN1 9QJ\\
}
\date{Accepted;
      Received;
      in original form 1999 Jan 31}

\pagerange{\pageref{firstpage}--\pageref{lastpage}}
\pubyear{1999}
\volume{}

\label{firstpage}

\maketitle

%%% ----------------------------------------------------------------------

\begin{abstract}

We present the first results of our HST WFPC2 F814W snapshot imaging
survey, targeting virtually
all sub-mJy decimetric radio-selected star-forming galaxies.
The radio
selection at $\sim1$ GHz is free from extinction
effects and the radio luminosities are largely
unaffected by AGN contamination,
making these galaxies ideal tracers of the cosmic star
formation history.
A sub-sample of $4$ targets is presented here, selected at $1.4$ GHz
from the spectroscopically homogenous and complete samples of Benn et
al. (1993) and
Hopkins et al. (1999). The redshifts are confined to a narrow range
around $z\sim0.2$, to avoid differential evolution,
with a radio luminosity close to $L_*$ where the galaxies
dominate the comoving volume-averaged star formation rate.
We find clearly disturbed morphologies
resembling those of ultraluminous infrared galaxies, indicating
that galaxy interactions may be the dominant mechanism for triggering star
formation at these epochs.
The morphologies are
also clearly different from coeval quasars and radiogalaxies, 
as found in star-forming galaxies selected at other wavelengths.
This may
prove challenging for models which propose direct causal links between AGN
evolution and the cosmic star formation history at these epochs.
The asymmetries are typically much larger than seen in the CFRS at
similar redshifts, optical luminosities and H$\alpha$-derived star
formation rates, indicating the possible existence of an
obscuration-related morphological bias in such samples.

\end{abstract}

\begin{keywords}
cosmology: observations -
galaxies:$\>$formation -
infrared: galaxies - surveys - galaxies: evolution -
galaxies: star-burst

\end{keywords}
\maketitle

\section{Introduction}

\begin{figure*}
\centering
%\vbox to2.5in{\rule{0pt}{2in}}
%\special{psfile=GAL-011225-454700.ps voffset=-310 hoffset=0
%vscale=80 hscale=80 angle=0}
  \ForceWidth{4.5in}
%  \BoxedEPSF{GAL-011225-454700.ps}
  \vspace*{-7cm}
  \hSlide{-1cm}
  \BoxedEPSF{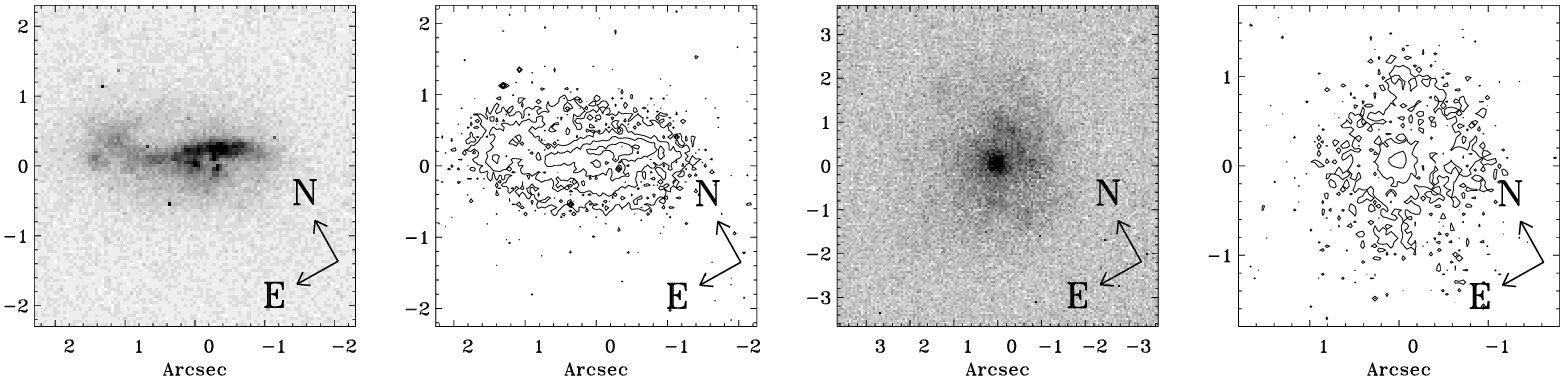}
  \ForceWidth{4.5in}
  \vspace*{-17.5cm}
  \hSlide{-1cm}
  \BoxedEPSF{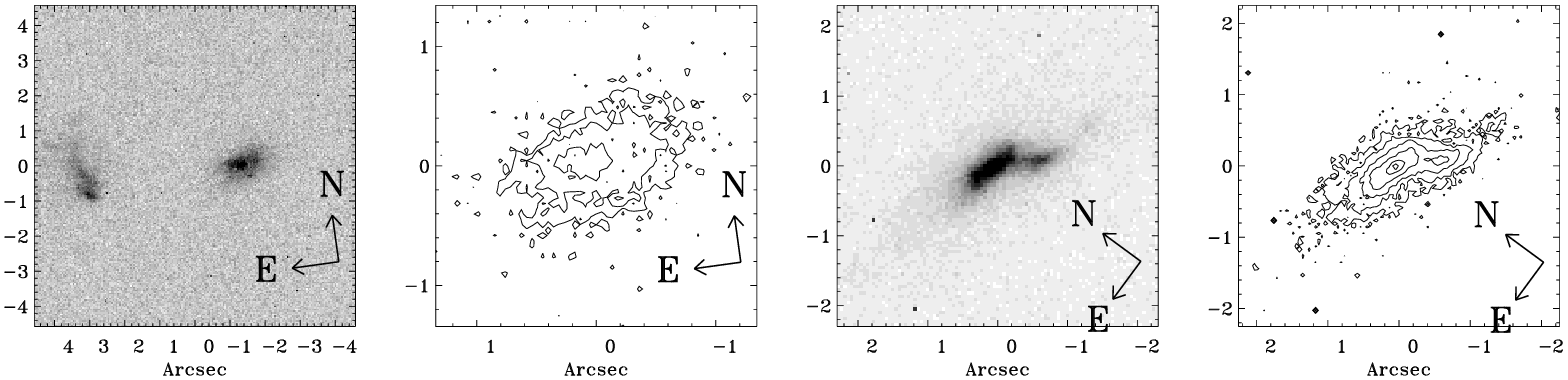}
  \vspace*{1cm}
\caption{
\label{fig:images}
HST F814W snapshots of the  galaxies
[CM84]144 (top left),  [CM84]074 (top right), Phoenix-Deep-29 (bottom
left) and Phoenix-Deep-96 (bottom right).
Greyscale is linear, and contours are spaced in
factors of $2$. 
%The lower right panel covers only a subset of the
%lower left image, while the areal coverages of the upper right and
%upper left panels are the same.
}
\end{figure*}

%\begin{figure*}
%\centering
%%\vbox to2.5in{\rule{0pt}{2in}}
%%\special{psfile=GAL-011225-454700.ps voffset=-310 hoffset=0
%%vscale=80 hscale=80 angle=0}
%  \ForceWidth{6in}
%%  \BoxedEPSF{GAL-011225-454700.ps}
%  \BoxedEPSF{snaps_fig1.ps}
%\caption{
%\label{fig:benn}
%HST F814W snapshots of the  galaxies
%%Coadd of $2\times 400$ second F814W exposures of the galaxies
%[CM84]144 (top) and [CM84]074 (bottom), both of which are members of
%the Benn et al. (1993) sample. Greyscale is linear, and contours are spaced in
%factors of $2$. The lower right panel covers only a subset of the
%lower left image, while the areal coverages of the upper right and
%upper left panels are the same.
%}
%\end{figure*}
%
%
%\begin{figure*}
%\centering
%  \ForceWidth{6in}
%%\vbox to3.3in{\rule{0pt}{2in}}
%%\special{psfile=GAL-011043-455122.ps voffset=-300 hoffset=0
%%vscale=80 hscale=80 angle=0}
%%  \BoxedEPSF{GAL-011043-455122.ps}
%  \BoxedEPSF{snaps_fig2.ps}
%\caption{
%\label{fig:phoenix}
%HST F814W snapshots of the galaxies
%Phoenix-Deep-29 (top) and Phoenix-Deep-96 (bottom).
%%Coadd of $2\times 400$ second F814W exposures of the galaxies
%Greyscale is linear, and contours are spaced in
%factors of $2$. The upper right panel covers only a subset of the
%upper left image, while the areal coverages of the lower right and
%lower left panels are the same.
%}
%\end{figure*}

\begin{table}
\begin{tabular}{lll}
Redshift & Evolution & $L_*$ $1.4$ GHz Flux \\
\hline
0.2 & $(1+z)^3$   & $0.264$ mJy\\
0.2 &      none   & $0.153$ mJy\\
0.3 & $(1+z)^3$   & $0.145$ mJy\\
0.3 &      none   & $0.066$ mJy\\
\hline
\end{tabular}
\caption{\label{tab:lstar}Flux of the break in the $1.4$ GHz
luminosity function ($L_*$) under various luminosity
evolution assumptions. Uses the
luminosity function from Serjeant et al. 1998, with an $\Omega=1$,
$\Lambda=0$ world model. Very similar results are derived from the
Condon 1992 luminosity function.
For a Hubble constant of $H_0=50$ km s$^{-1}$ Mpc$^{-1}$
the characteristic luminosity
lies at $\sim2.8\times10^{22}$ W Hz$^{-1}$.
The fluxes themselves are $H_0$-independent.
}
\end{table}

The study of the observational
constraints on the cosmic star formation history is currently among the
most active fields in observational cosmology.
The most widely used tracer of the comoving volume-averaged star
formation rate (SFR) is the UV luminosity density (e.g. Madau et
al. 1996, Steidel et al. 1996, 1999), found to
peak at $z\sim1-2$. However, little is known about the history of star
formation in the Universe beyond its global average (e.g. Abraham et
al. 1999). 
%In this paper, we present the first results from an
%on-going program with the Hubble Space Telescope (HST) to study
%morphological types of the star-forming galaxies, dominating the comoving
%SFR, free from obscuration-dependent biases.
In this paper we present the first results from an on-going program
with Hubble Space Telescope (HST) 
to study the morphology of starforming galaxies, using a sample
of galaxies which dominate the comoving SFR and 
which were selected in a manner free from obscuration biases.

\begin{table*}
\begin{tabular}{lllllllllll}
Name 
%& 
%  Date     
      & RA           & Dec        & Mag & $z$ & $S_{1.4}$ & I$_{814}$       & $M_B$ & SFR        & SFR        & A\\
%    &  
%  Observed 
    &   (J2000)    &  (J2000)   &     &     & (mJy)     &  & $(AB)$& ($1.4$)& (H$\alpha$)    \\
\hline
{[CM84]144}$^{2,3,4}$ 
%& 
%       GAL-085547+170228 
%    07/01/2000    
                   & $08$ $55$ 
                   $47.60$ & $+17$ $02$ $28.4$ 
                   & B $= 18.7 $ & $0.2253$ & $0.29$ & $18.3$ & $-21.2$ & $116.3$ & $12.9$ & $0.47$ \\ % visit 1
%{65W115} & GAL-084841+445109        & $08^{\rm h}$ $48^{\rm m}$ $41.488^{\rm s}$ &  $+44^\circ$ $51'$ $9.87''  $ & V $= 22   $ & $0.1937$ & $0.15$ &3,4,6\\ % visit 2
%{65W277} & GAL-085029+450450        & $08^{\rm h}$ $50^{\rm m}$ $29.813^{\rm s}$ & $+45^\circ$ $04'$ $50.70'' $ & V $= 18.4 $ & $0.1909$ & $0.23$ &3,4,6\\ % visit 3
%{[MC85]063} & GAL-130251+301049     & $13^{\rm h}$ $02^{\rm m}$ $51.229^{\rm s} $ & $+30^\circ$ $10'$ $49.99'' $ & B $= 19.5 $ & $0.1710$ & $0.24$ & 3,5,6\\ % visit 4
{[CM84]074}$^{2,3,4}$ 
%& 
%       GAL-085458+170346   
%    07/01/2000  
                   & $08$ $54$ $58.33$ & 
                   $+17$ $03$ $46.9$ 
                   & B $= 18.6 $ & $0.2279$ & $0.29$ & $18.7$ & $-20.8$ & $117.6$ & $8.34$ & $0.112$ \\ % visit 5
%{65W137} & GAL-084853+444626        & $08^{\rm h}$ $48^{\rm m}$ $53.433^{\rm s} $ & $+44^\circ$ $46'$ $26.41''$  & V $= 20.4 $ & $0.28$   & $0.11$ & 3,4,6\\ % visit 6
%{[CM84]016} & GAL-085415+170137     & $08^{\rm h}$ $54^{\rm m}$ $15.6612^{\rm s}$ & $+17^\circ$ $01'$ $37.338''$ & B $= 19.2 $ & $0.1835$ & $0.25$ & 2,3,6\\ % visit 7
Phoenix-Deep-29$^{1,5,6}$ 
%& 
%       GAL-011043-455122 
%    01/09/1999 
                 & $01$ $10$ $43.88$ 
                 & $-45$ $51$ $22.7$ 
                 & R $= 20.2$ & $0.210 $ & $0.198$ & $19.7$ & $-19.6$ & $52.5$ & $1.1$ & $0.185$ \\ % visit 8
%Phoenix-Deep-52 & GAL-011126-453416 & $01^{\rm h}$ $11^{\rm m}$ $26.906^{\rm s} $ & $-45^\circ$ $34'$ $16.26'' $ & R $= 18.57$ & $0.190 $ & $0.141$ & 1,7,8\\ % visit 9
Phoenix-Deep-96$^{1,5,6}$ 
%& 
%       GAL-011225-454700 
%    18/07/1999  
                & $01$ $12$ $25.21$ 
                & $-45$ $47$ $0.1$ 
                & R $= 18.4$ & $0.236 $ & $0.225$ & $18.3$ & $-21.2$ & $75.9$ & $31.0$ & $0.317$\\ % visit 10
\hline
\end{tabular}
\caption{\label{tab:sample}HST $z\sim0.2$ sub-mJy star-forming snapshot
sample observed to date. All were observed
for $2\times400$ seconds with the PC in filter F814W.
References are 
 (1) Georgakakis et al. 2000, 
 (2) Condon \& Mitchell 1984, 
 (3) Rowan-Robinson et al. 1993,
%(4) Oort et al. 1987, 
%(5) Mitchell \& Condon 1985, 
 (4) Benn et al. 1993,
 (5) Hopkins et al. 1998, 
 (6) Hopkins et al. 1999. 
%(7) Oort et al. 1987,
%(8) Mitchell \& Condon 1985
Star formation rates ($1.4$ GHz and H$\alpha$)  are in
$M_\odot$ yr$^{-1}$,  and $A$ are the asymmetries. 
For comparison, the $z=0.2$ ($z=0.3$) radio
$L_*$ assuming $(1+z)^3$ luminosity evolution corresponds
to $63.4$ M$_\odot$ yr$^{-1}$ ($80.6$ M$_\odot$ yr$^{-1}$). The Tresse
\& Maddox (1997) H$\alpha$ $L_*$ at $z\sim0.2$ corresponds to $22.4$
M$_\odot$ yr$^{-1}$.
}
\end{table*}

An important caveat to the SFR constraints is that the UV luminosity of
star-forming galaxies is dominated by the lowest-extinction regions. This
leads to extreme sensitivity to obscuration corrections (e.g. Pettini et  
al. 1998, Meurer et
al. 1997), large enough to eliminate the evidence for a
redshift cut-off in the SFR.
%An important caveat to the SFR constraints is that
%the UV luminosity of star-forming galaxies is dominated by
%the lowest-extinction regions, leading to extreme sensitivity to
%obscuration corrections (e.g. Pettini et al. 1998, Meurer et
%al. 1997), large enough to eliminate the evidence for a
%redshift cut-off in the SFR.
Indeed, SCUBA observations of the Hubble Deep Field
(HDF) (Hughes et al. 1998) detected $5$ ultraluminous
galaxies at $z>2$ (the redshift constraints come mainly
from the radio-FIR correlation and not from the uncertain HDF IDs),
which together implied an obscured star formation rate at least as
large as the integrated
de-reddened UV-derived rate. Moreover, the H$\alpha$-derived star
formation rate from the Canada-France Redshift Survey (CFRS, Tresse \&
Maddox 1997) was a factor $\sim\times3$
larger than the UV estimate in the same sample, but comparable to
those derived from
the mid-IR for the CFRS (Flores et al. 1999) or the HDF
(Rowan-Robinson et al. 1997).

However, a serious problem is
that optical-UV light is strongly skewed to low extinction regions,
while the reverse is true for mid and far-IR selected samples,
leading to problematic obscuration-related effects in either case.
An unbiased technique for selecting star-forming galaxies is to sample at
decimetric radio wavelengths, where both obscured and unobscured star
forming galaxies contribute (Condon 1992).
Several groups (including ourselves)  have begun to
exploit this technique to trace the comic star formation history,
and to study the obscuration effects in star forming galaxies
(e.g. Cram et al. 1998, Serjeant et al. 1998, Oliver et al. 1998,
Haarsma \& Partridge 1998, Cram 1998, etc).
%It has been shown
%that the sub-mJy radio population, selected at 1.4 GHz, is dominated by
%star-forming galaxies, with the contributions from AGNs or bright
%ellipticals reducing by $50\%$ (Hopkins et al 1999).

We have embarked on a programme to image almost all star-forming
galaxies
selected at decimetric radio wavelengths, using the WFPC2 in snapshot
mode on the HST.
In this paper we carry out a morphological study of
a small subset of these galaxies at $z\sim0.2$, from our first
cycle $8$ observations in this programme.
(Although these are at low redshifts by many standards, they are {\it not}
local, with the volume-averaged SFR being a factor of $\sim\times2$
higher than
in the local Universe for $(1+z)^3$ luminosity evolution models (Table
\ref{tab:lstar}).)
%Note that although these are low redshifts by many
%standards, these galaxies are {\it not} local: the volume-averaged
%SFR is $\sim \times 2$ higher than in the local Universe
%for $(1+z)^3$ luminosity evolution.)
%using $I_{\rm F814W}$-band HST WFPC2 snapshot observations.
The galaxies in this sample are selected so that they dominate the
(radio-derived) star formation history of the Universe at this epoch,
i.e. close to the $L_*$ characteristic luminosity in the
radio luminosity function. 

The sample selection is discussed in Section 2, which also
presents the HST
observations and data reduction. These data are then
analysed in Section 3, followed by a discussion of the results in
Section 4.

\begin{figure}
  \ForceWidth{4in}
  \hSlide{-1cm}
%\vbox to3.3in{\rule{0pt}{2in}}
%\special{psfile=GAL-011043-455122.ps voffset=-300 hoffset=0
%vscale=80 hscale=80 angle=0}
\vspace*{-1cm}
  \BoxedEPSF{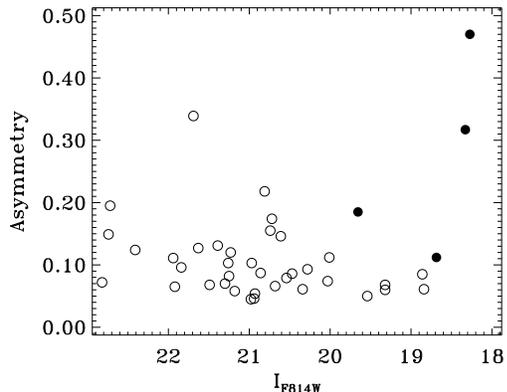}
\vspace*{-1cm}
\caption{
\label{fig:ai}
Asymmetry parameters for the radio-selected
sub-mJy star-forming galaxies (filled symbols)
compared to the CFRS/LDSS galaxies with $0.17\le z\le 0.3$ (open symbols).
}
\end{figure}

\section{Sample selection and observations}\label{sec:method}
The following criteria are considered in selecting the sample for HST
observations:

\begin{enumerate}
\item The objects are sub-mJy radio sources, selected at 1.4 GHz
(Benn et al (1993); Georgakakis et al (2000)). This
avoids obscuration-related selection biases.
%A significant fraction of sub-mJy 1.4 GHz
%source counts is produced by star-forming galaxies at intermediate
%redshift (e.g. Benn et al. 1993). 
%Furthermore, the d
Decimetric radio
fluxes are much less affected by AGN contribution
than shorter wavelength observations
(compare e.g. the ongoing contoversy over the AGN fraction in
higher frequency sample with Hammer et al. (1995) finding
$\sim 50\%$ AGN fraction at $5-8$ GHz while Windhorst et al. (1995)
find only $\sim 5\%$).

\item Spectroscopic data are used to select 
%%those sub-mJy sources that are
emission-line star-forming galaxies.

\item The galaxies must not lie close to bright stars to avoid
HST roll angle constraints.

\item Galaxies selected to lie close to the $1.4$GHz $L_*$, which
dominate the volume averaged SFR (Table \ref{tab:lstar}).

\item Galaxies in a narrow redshift range around
$z\sim0.2$ to avoid differential evolution.

\end{enumerate}

Our primary sample satisfies (i), (ii) and (iii), 
%using the radio surveys
%in Benn et al (1993) and Georgakakis et al (1999). 
and is the 
subject of our $150$ snapshots in cycles $8$ and $9$. Ten galaxies
also satisfy conditions (iv) and (v); these are
the cycle $8$ targets. 
%A total of $10$ galaxies satisfied these criteria, 
The HST observations for
%a sub-set of four galaxies in this sample 
four of these
have now been completed.
We selected the F814W (wide I band) filter. 
At this wavelength, the light
is mostly dominated by old stellar population, so gives a more 
accurate measure of distortion in the underlying gravitational
potential than e.g. UV observations.
Details of the current sub-sample are presented in Table \ref{tab:sample}.

%\section{HST Observations and Data Reduction}\label{sec:reduction}
%The HST observations were carried out in cycle $8$. 
%Table
%\ref{tab:log} presents the observing log.

The data were reduced using the Interactive Data Language (IDL),
starting from the automatic pipeline products. Cosmic rays were
identified as $>3\sigma$ differences between
frames. For sky subtraction, we estimated the modal value of
the underlying counts distribution using iterative fits to the
readout histograms. The final HST images for the four sub-mJy radio
sources
are presented in Figure \ref{fig:images}.

\section{Results}

%In this Section we first discuss the individual morphologies,
%then quantify the asymmetries.
%\subsection{Details of individual objects}\label{sec:objects}

%\subsubsection{[CM84] 144}
%This galaxy 
[CM84] 144 is extremely disturbed, with a tidal tail
extending $\sim1.5''$ North-East of the nucleus. Several secondary
nuclei are also apparent,
reminiscent of HST WFPC2
F814W imaging of ultra-luminous
infrared galaxies (e.g. Borne et al. 2000) such as Arp 220
(Borne \& Lucas 1997).

%\subsubsection{[CM84] 074}
%This galaxy 
[CM84] 074 is only mildly asymmetric. There are hints of face-on
spiral structure, and an excess flux $\sim0.5''$ West of the nucleus.
There are two companion galaxies a few arcseconds to the West, also
visible in the digitised sky survey. (We reject these as
identifications of the radio source based on the ID magnitude quoted
in Benn et al. 1993.)

%\subsubsection{Phoenix-Deep-29}

Galaxy Phoenix-Deep-29 is again not as clearly disturbed.
However, the inner isophotes
are clearly offset in position and (tentatively) in orientation from
the outer isophotes. This galaxy also shows
signs of
interaction with a companion $5''$ to the East.
%(The likelihood of such a
%galaxy being within this distance is $<BLAH$, on the basis of the HST
%MDS I-band counts.)
There is a clear tidal tail from the companion extending a few
arcseconds North.

%\subsubsection{Phoenix-Deep-96}
Galaxy Phoenix-Deep-96 shows clear signs of morphological
disturbance, with the structure dominated by a bright central bar-like
feature.
There are also hints of
multiple nuclei inside the bar, and there is a secondary peak
$\sim0.5''$ South-West of the bar.
There are however
no other clearly associated companion galaxies visible in the HST image.
% indicating that a significant
%merger has already taken place (unless, rather improbably, the
%companion lies along the same line of sight as the star-forming).

%There is some evidence for distorted isophotes in both the
%sub-mJy star-forming and its companion.

%Discussion of radio-optical alignments if any?

%\subsection{Quantitative measurement of flux asymmetry}\label{sec:asymmetry}
To quantify the morphological disturbance, we follow the procedure
adopted by Brinchmann et al. (1998) and others, by
rotating each image through $\pi$ and subtracting it from
the original
image. Normalising this to the galaxy flux yields the fractional
asymmetric flux:
\begin{equation}
A = \frac{\Sigma | G_{ij} - G'_{ij} |}{\Sigma G_{ij}} - k
\end{equation}
where $G$ ($G'$) is the rotated (unrotated) image, $k$ is a
correction for systematics such as sky gradients, and the sum is
performed over pixels $ij$.
%, and the sums are
%performed over pixels above a given surface brightness threshold.
The values of $A$ are fairly insensitive to the choice of aperture used
for the self-subtraction, provided all pixels at
$\stackrel{>}{_\sim}1.5\sigma$ are included in the subimage.
The magnitude of $k$
can be estimated by applying the technique to blank sky regions of the
same size. The errors in $A$ are in practice dominated by
the uncertainty in $k$: we find typically $\Delta A\simeq 0.05$.
A further advantage of this statistic is that the
$I$-selected Canada-France Redshift Survey (CFRS) and $b_{\rm
J}$-selected LDSS surveys have extensive published HST morphologies
quantified using the same statistic, yielding valuable control
samples. The asymmetry measures
for the sub-mJy star-forming galaxies, quantified as discussed above,
are listed in Table \ref{tab:sample}.

%\subsection{Fourier analysis}\label{sec:fourier}

\section{Discussion and conclusions}\label{sec:discussion}
In Figure \ref{fig:ai} we plot the asymmetry statistics of our
galaxies, and
compare them to a control sample of coeval galaxies from the CFRS
sample (Brinchmann et al. 1998). The control extends to fainter
$I_{\rm F814W}$ fluxes than our star-forming galaxy sample, though
there is some
overlap. Remarkably, our galaxies are far more asymmetric than the
control sample of an (ostensibly) identical population of
optically-selected star-forming
galaxies of the same $I_{\rm F814W}$. 
%This is also apparent by visual
%inspection of Figures \ref{fig:benn} and
%\ref{fig:phoenix}, since several of our sample are quite
%clearly highly disturbed.

\begin{figure}
  \ForceWidth{4in}
  \hSlide{-1cm}
\vspace*{-1.5cm}
%\vbox to3.3in{\rule{0pt}{2in}}
%\special{psfile=GAL-011043-455122.ps voffset=-300 hoffset=0
%vscale=80 hscale=80 angle=0}
  \BoxedEPSF{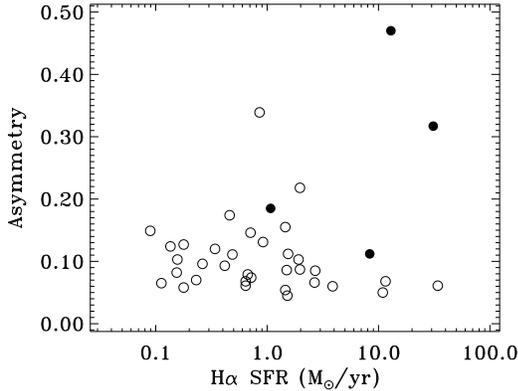}
\vspace*{-1cm}
\caption{
\label{fig:asfr}
Asymmetry parameter plotted against the
estimate of the H$\alpha$ star formation rate discussed in the text.
Symbols as in Figure \ref{fig:ai}.
}
\end{figure}

The optical control must be well-matched with the radio sample
in as many {\it optical} properties as possible.
The radio sample is more asymmetric
at a fixed optical luminosity, but is this also true at a fixed
optically-derived SFR?
I.e., could the morphological differences between the radio sample and
the optically-selected control be instead because they are not
well-matched in optically-derived star formation rates?
To address this we must compare the H$\alpha$
luminosities of our radio sample with those of the control.
The data for the radio samples are presented in
Benn et al (1993) and Georgakakis et al (2000).
The individual H$\alpha$ luminosities of the CFRS galaxies are not
published, but
Tresse \& Maddox (1998) report a tight correlation between H$\alpha$
luminosities and $M_{\rm B}(AB)$ absolute magnitudes. In Figure
\ref{fig:asfr} we
use this
relation to estimate the H$\alpha$-derived star formation
rates, and compare the optical control sample with our sub-mJy
star-forming galaxies.
We adopt the following conversion between H$\alpha$ luminosities and
star-formation rates (SFRs):
\begin{equation}
{\rm SFR} ({\rm H}\alpha) =
  L({\rm H}\alpha)/(1.41\times 10^{34} {\rm W})
\end{equation}
which is derived assuming Salpeter IMF between
$0.1$ $M_\odot$ and $125$ $M_\odot$
(see Serjeant et al. 1998, Oliver et al. 1998).
We de-redden our H$\alpha$ fluxes by $A_{\rm V}=1$ for
consistency with optically-selected samples.
(Tresse \& Maddox 1998 note that the H$\alpha$ luminosity function is
essentially unchanged if assuming this $A_{\rm V}$ throughout their
sample, instead of using individual Balmer decrements.)
%Note that it is important to treat the extinction corrections of the
%control sample and the radio sample consistently
There is a more substantial overlap of the
H$\alpha$-derived SFR with the control sample, and it is again clear
that at a {\it fixed} H$\alpha$ SFR the radio-selected sample is
significantly more disturbed than the optically-selected sample of
star forming galaxies.

In summary, the radio-selected sources are typically more
morphologically
disturbed than the optically-selected control, well-matched in
redshift, optical luminosity and/or H$\alpha$ SFR.
%It would appear therefore that the
%{\it most} disturbed morphologies are under-represented in optical
%samples of star forming galaxies.
%At a fixed redshift, optical luminosity and/or H$\alpha$ SFR the
%radio-selected sample is systematically more asymmetrical.
Why this might be the case is hinted
in Table \ref{tab:sample}, where we compare the star formation rates of
our galaxies estimated from H$\alpha$ and the $1.4$GHz radio
luminosity. To convert decimetric radio luminosities to SFRs we use
\begin{equation}
{\rm SFR} (1.4 {\rm GHz}) =
  L_{1.4}/(7.63\times 10^{20} {\rm W~Hz}^{-1})
\end{equation}
appropriate for the same IMF as above
(see Oliver et al. 1998, Serjeant
et al. 1998).
Our H$\alpha$ SFRs are comparable in most cases
to $L_*$ in the H$\alpha$ luminosity
function (Tresse \& Maddox 1997). However the radio-derived star
formation rates are much larger (though still comparable with the
radio $L_*$, Table \ref{tab:lstar}). This indicates that a large fraction
of the star formation in these objects occurs in regions with
$A_{\rm V}\gg1$, since the observed H$\alpha$ luminosities will be
dominated by regions with $A_{\rm
V}\stackrel{<}{_\sim}1$. Indeed if the dust is well-mixed with the
H$\alpha$ emitting gas, the different optical depths for H$\alpha$ and
H$\beta$ ensure that $A_{\rm V}=1.1$ would be derived for a simple
screen Balmer decrement model {\it regardless} of the true extinction
to the rear of the cloud.
We therefore suggest that optically-selected samples may
under-represent disturbed galaxies with large amounts of obscured star
formation, in any SFR-weighted quantity.

The galaxy morphologies of the sub-mJy sources are also markedly
different from coeval AGN.
McLure el al. (1999) and McLeod \& Reike (1995) have
found giant ellipticals hosting both radio-loud and radio-quiet
quasars at these
redshifts, and radiogalaxies are also early type galaxies at these
epochs. The EMSS-selected sample of Schade et al. (1999) at
$z\le0.15$ also
finds no evidence for strong interaction/merger activity for any AGN
in the sample.
This may pose
problems for models which propose causal links between the cosmic
evolution of AGN and the cosmic star formation history. For example,
a model in which both are driven by
similar merger events must also provide a
mechanism for delaying the onset of central engine fuelling, to account
for the (relatively) more relaxed AGN host galaxy potentials.
This is related to the more fundamental and unsolved
problem of AGN evolution:
how to drive the gas fuel down $\sim 5$ orders of magnitude in radius.

We can also compare our sample with the ultraluminous
galaxy samples of e.g. Borne et al. (2000), who also used
HST WFPC2 F814W in snapshot mode. These authors found highly
disturbed systems often
in moderately rich environments, which they used
to argue for an evolutionary sequence involving compact groups
and ultraluminous galaxies. Although asymmetry statistics have
not been calculated for the ultraluminous galaxies, our targets
appear qualitatively less disturbed and the environments possibly
less rich.
(Ideally both ultraluminous galaxies and
radio-selected star-forming galaxies should also
have their environments quantified
with for example the $B_{\rm gg}$ statistic; e.g.
Yee \& Green 1984, Hill \& Lilly 1991, Wold et al. 2000)
This may suggest a link between the level
of star formation activity after interactions and the richness
of the environment, perhaps via the number or the nature of the
interaction events.
% (Hopkins et al. 2000).

Although our current sample is small, the asymmetric morphologies
in our $\sim L_*$ targets suggest that galaxy interactions play a
major role
in the evolution of the star formation and metal production rates
in the low-$z$ Universe. Further papers in this
series will present results
for larger samples of sub-mJy star-forming galaxies. However, it is
worth keeping in mind that
the formation and evolution of galaxies cannot be characterised
by a single parameter, such as the volume-averaged star formation
rate. In expanding on this simple first-order description,
the obscuration-independent selection of galaxies will be essential,
as will co-ordinated multi-wavelength follow-up (e.g. ELAIS, Oliver et
al. 2000). 
%The European Large Area ISO Survey (ELAIS, Oliver et al. 1999)
%and its follow-ups are ideal in this
%regard, as they combine sub-mJy decimetric radio data with
%extensive $7-175\mu$m mid-far-IR data, $U$-band imaging and (in
%selected areas) $850\mu$m and hard X-ray survey data. The ELAIS
%survey also covers enough comoving volume to eliminate
%large scale structure fluctuations in its volume-averaged
%quantities, a problem which has plagued every estimate of the
%$z\sim0.2$ comoving SFR. In the longer term the FIRST and SIRTF
%missions will survey $\sim 10-100$ square degrees at $60-500\mu$m,
%and a $\mu$Jy decimetric ATCA radio survey covering a comparable area
%is also under discussion.
%These will extend the redshift range from $z\stackrel{<}{_\sim}1.5$
%in ELAIS to $z\sim4$.
%With larger and higher redshift
%samples of sub-mJy star-forming galaxies with complementary
%multi-wavelength survey data, it will
%be possible to investigate the star formation history as a
%function of (for example) Hubble type, galaxy environment, or
%dust content.
%With the cosmological parameter estimates from future
%microwave background missions it will also be possible to estimate the
%evolving bias parameter of star-forming galaxies.

%The future: the Madau diagram as a function of Hubble type, from
%dust-independent samples.

%Comparison with AGN, ULIRGs and other star-forming galaxies (HDF or
%HST MDS for example?)

\section*{Acknowledgements}
It is a pleasure to thank Patricia Royle for her help in the preparation
of these observations.
Based on observations with the NASA/ESA Hubble
Space Telescope, obtained at the Space Telescope
Science Institute, which is operated by the
Association
of Universities for Research in Astronomy, Inc. under
NASA contract No. NAS5-26555.
This work was
supported by PPARC (grant
number GR/K98728) and by the EC TMR Network programme
(FMRX-CT96-0068).

% ------------------------------------------------------------------------
%\bibliographystyle{mn}
%\bibliography{}
\end{document}